\documentclass[twocolumn,twoside,slac_two,floatfix]{revtex4}
\usepackage{graphicx}
\usepackage{fancyhdr}
\newcommand{\aap}{A\&A}
\newcommand{\apjs}{ApJ}

\begin{document}

\title{H I spin temperature in the Fermi-LAT era}

%

\author{G. J{\'o}hannesson}
\affiliation{Hansen Experimental Physics Laboratory, Stanford University, Stanford, CA 94305, USA}
\author{S. Digel}
\affiliation{Stanford Linear Accelerator Center, Menlo Park, CA 93025, USA}
\author{I. Moskalenko}
\affiliation{Hansen Experimental Physics Laboratory, Stanford University, Stanford, CA 94305, USA}
\author{on behalf of the Fermi Large Area Telescope Collaboration}

\begin{abstract}
The diffuse high-energy gamma-ray emission of the Milky Way arises from
interactions of cosmic-rays (CRs) with interstellar gas and radiation
field in the Galaxy.  The neutral hydrogen (H I) gas component is by far
the most massive and broadly distributed component of the interstellar
medium.  Using the 21-cm emission line from the hyperfine structure transition of
atomic hydrogen it is possible to determine the column density of H I if
the spin temperature ($T_S$) of the emitting gas is known.  Studies of
diffuse gamma-ray emission have generally relied on the assumption of a
fixed, constant spin temperature for all H I in the Milky Way.
Unfortunately, observations of H I in absorption against bright
background sources has shown it to vary greatly with location in the
Milky Way.  We will discuss methods for better handling of spin
temperatures for Galactic diffuse emission modeling using the Fermi-LAT data
and direct observation of
the spin temperature using H I absorption.
\end{abstract}

\maketitle

\thispagestyle{fancy}


\section{Introduction}
The diffuse Galactic emission (DGE) arises from
interactions of cosmic-rays (CRs) with interstellar gas and radiation
field in the Galaxy.  Due to the smooth nature of the
interstellar radiation field and the CR flux after propagation, the fine structure of the DGE is
determined by the structure of the interstellar gas.  Getting the
distribution of the interstellar gas correct is therefore crucial when
modeling the DGE.

It is generally assumed that Galactic CRs are accelerated in interstellar
shocks and then propagate throughout the Galaxy \citep[see e.g.][for a
recent review.]{Strong2007}.  In 
this paper, CR propagation and corresponding diffuse emission is
calculated using the GALPROP code \citep[see][and references
within.]{Strong2004}.  We use the so-called
conventional GALPROP model \citep{Strong2004}, where the CR injection spectra and the
diffusion parameters are chosen such that the CR flux agrees
with the locally observed one after propagation.  The gas distribution is
given as Galacto-centric annuli and the diffuse emission is calculated
for those same annuli.  The distribution of H I is
determined from the 21-cm LAB line survey \citep{Kalberla2005} while
distribution of molecular hydrogen, H$_2$,
is found using the CO ($J=1\rightarrow0$) survey of \cite{Dame2001}
assuming $N_{\rm{H}_2} = X_{CO}(R) W_{CO}$.

\begin{figure}
	\begin{center}
		\includegraphics[width=75mm]{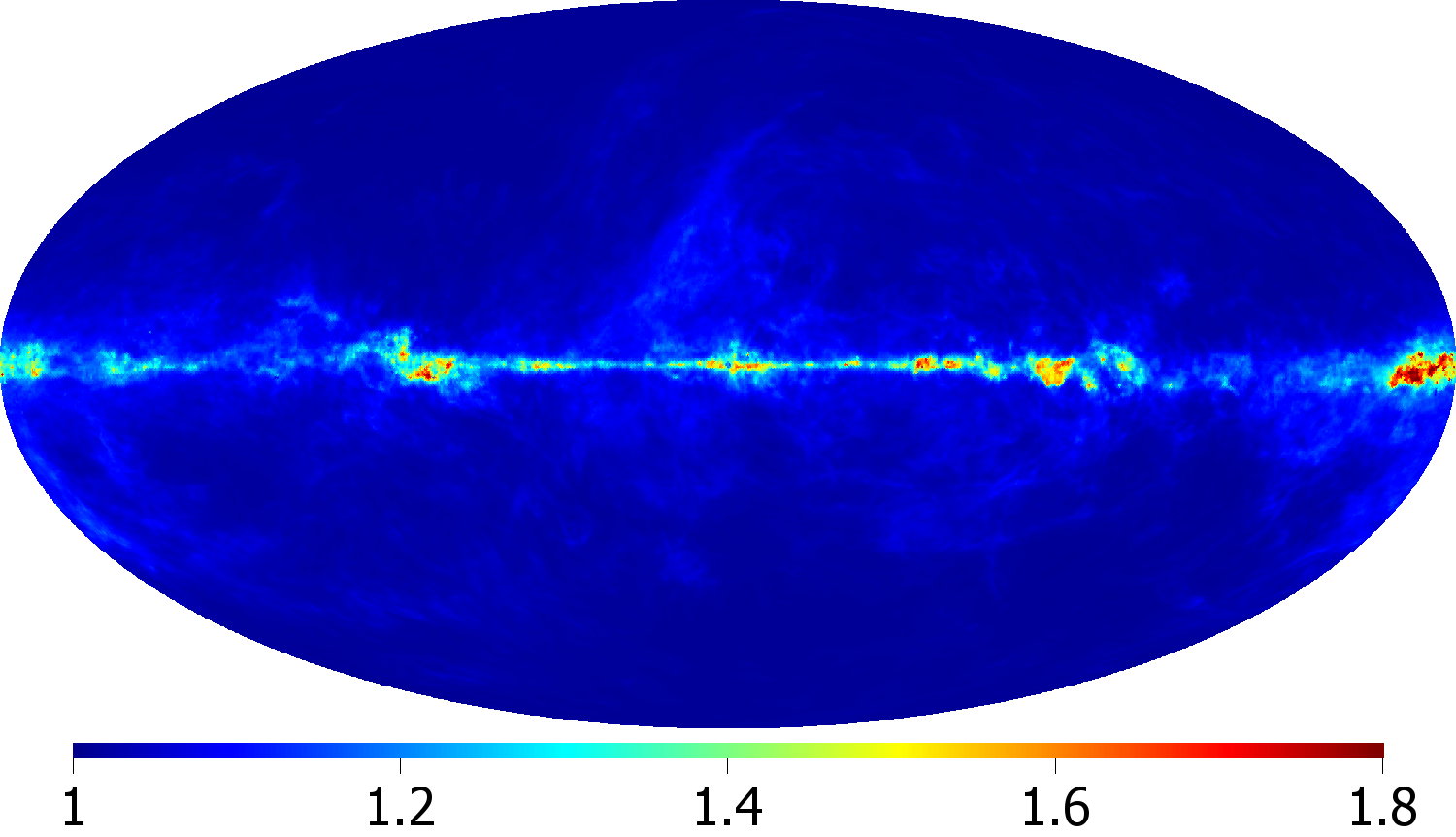}
	\end{center}
	\caption{The ratio $N_{H I}(125\,
\rm{K})/N_{H I}(200\,\rm{K})$ in
Galactic coordinates.  The figure clearly shows the non-linearity of the
correction that can be as high as a factor of 2 in this case.}
	\label{fig:TsRatio}
\end{figure}

While converting observations of the 21-cm H I line to column density is
theoretically possible, it is not practically feasible.  To correctly
account for the optical depth of the emitting H I gas, one must know its
spin temperature, $T_S$ \citep[see e.g.][]{Kulkarni1988}.
Under the assumption of a constant $T_S$ along the line of sight, the
column density of H I can be calculated from the observed  brightness
temperature $T$ using
\begin{equation}
	N_{H I}(v,T_S) = -\log\left(1-\frac{T}{T_S-T_{bg}}\right) T_S C,
	\label{eq:OpticalDepthCorrection}
\end{equation}
where $T_{bg}$ is the background continuum temperature and $C =
1.83\times10^{18}$ cm$^{-2}$
K (km/s)$^{-1}$.  The assumption of a constant $T_S$ along the line of
sight is known to be wrong for many directions in the Galaxy \citep[see
e.g.][]{Dickey2009}.  The $T_S$ values derived in this paper are
therefore only a global average and should not be taken at face value.
Figure~\ref{fig:TsRatio} shows how changing $T_S$
affects $N_{H I}$ in a non-linear way, mainly affecting areas with $T$
close to $T_S$ in the Galactic plane.  This figure was created under the
assumption of a fixed $T_S$ for the whole Galaxy that is known to be
wrong but has been used for DGE analysis from the days of
COS-B \citep{Strong1985}.  Note that for
equation~(\ref{eq:OpticalDepthCorrection}) to be valid the condition
$T_S > T+T_{bg}$ must hold.  When generating the gas annuli, this
condition is forced by clipping the value of $T$.

While the assumption of a constant spin temperature $T_S = 125\,\rm{K}$ for the whole
Galaxy may have been sufficient for older instrument, it is no longer
acceptable for a new generation experiment like Fermi-LAT
\citep{Atwood2009}.  This has
been partially explored for the outer Galaxy in \citet{2ndQuadrant}.  In this
paper we will show a better assumption for $T_S$ can be easily found and
also show that direct observations of $T_S$ using absorption measurement
of bright radio sources are needed for accurate DGE modeling.

\section{Method}

We assume the source distribution of CR nuclei and electrons are the
same.  CR propagation is handled by GALPROP and we use the conventional
model so that after the propagation the CR spectra agree with local
observations.  The GALPROP diffuse emission is output in Galacto-centric
annuli, split up into different components corresponding to different
processes (bremsstrahlung, $\pi^0$-decay, and inverse Compton scattering).  To
allow for radial variations in CR intensity we perform a full sky maximum
likelihood fit,
preserving the spectral shape of each component.  We allow for one
global normalization factor for the electron to proton ratio.
Additionally, we also allow for radial variation in the $X_{CO}$ factor.
This accounts for uncertainties in the CR source distribution and
$X_{CO}$ factor.

The maximum likelihood fits were performed on the whole sky using the
GaRDiAN package \citep{Abdo2009} after preparing the Fermi-LAT data with
the science tools.  We use the same dataset as \cite{Ackermann2009} that
has special cuts to reduce CR background contamination compared to the
standard event selection \citep{Atwood2009}.  In addition to the
DGE model, we also include all sources from the 1 year Fermi-LAT source
list \citep{Abdo2010} and an isotropic component to account
for EGB emission and particle contamination.  This fit is
performed for different assumptions of $T_S$ and a likelihood
ratio test is used to compare the quality of the fits.

\section{Results}

The simplest assumption is that of a constant $T_S$ for the whole Galaxy
and it deserves some attention
for historical reasons.  It will also serve as a baseline model for
comparison with other assumptions.  
To get an approximation for the best model, we scan $T_S$ from 110 K to
150 K in 5 K steps.  Our results show that $T_S=130\,\rm{K}$ gives the
maximum likelihood for this setup.

One of the problems with the constant global $T_S$ approximation, apart
from the fact that observations of the interstellar gas have shown it 
to be wrong, is that the
maximum observed brightness temperature in the LAB survey is $\sim$150 K
which is greater than our best fit global $T_S$.  This is solved by
clipping the observations when generating the gas annuli, which is not
an optimal solution.  A different possibility is to use  
the assumption
\begin{equation}
	T_S = \max(T_{S,\rm{min}}, T_{\rm{max}}+\Delta T_S).
	\label{eq:LinearTS}
\end{equation}
Here, $T_{\rm{max}}$ is the maximum observed brightness temperature for each line
of sight.  This ensures $T_S$ is always greater than $T$.
Scanning the values of 
$T_{S,\rm{min}}$ and $\Delta T_S$ with a step size of 10 K and 5 K,
respectively, gives us a maximum likelihood for 
 $T_{S,\rm{min}} = 110\,\rm{K}$ and $\Delta T_S=
10\,\rm{K}$.  While this assumption still does not account for the
complexity of the interstellar medium, the log likelihood ratio between the best fit linear
relation model and the best fit constant $T_S$ model is of the order of
1000, a significant change.


The most accurate $T_S$ estimates come from observations of H
I in absorption against bright radio sources.  We gathered over 500
lines of sight with observed $T_S$ from the literature
\citep{Strasser2004, Heiles2003, Dickey2009}.  This
covers about 0.2\% of the pixels in the LAB survey, allowing for accurate
column density estimates only in those pixels.  After taking our best
fit linear relation model and correcting the pixels with known $T_S$ the
fit was redone for the whole sky.  Note that we did not change the values of
$T_{S,\text{min}}$ and $\Delta T_S$.  The log likelihood ratio of --105
tells us
that this model is worse than the best fit linear relation.  This is not
unexpected, since the gamma rays are generated from CR interactions with
the gas and if the gas distribution is wrong, we won't get the correct CR
distribution from the fit.  To limit the
uncertainty involved with the linear relation assumption, we did another
fit, limiting ourselves to the region $-10^\circ < |b| < 10^\circ$,
$15^\circ < l < 165^\circ$ that
covers the observations made in the Canadian Galactic plane survey
(CGPS) where the density of $T_S$ observations is the highest and is large
enough to get a good fit to the LAT data.  The fit in this region
results in a log likelihood ratio of 28 indicating a statistically
significant improvement in the fit.  This is despite the observed $T_S$
lines of sight only covering 25\% of the fitted region and the values of
$T_{S,\rm{min}}$ and $\Delta T_S$ not being adjusted after correcting for known $T_S$ values.

\section{Discussion}

Our small exercise here has shown that for accurate DGE modeling we
need to know more about the distribution of gas in the Galaxy,
especially the H I distribution.  The standard constant $T_S$ assumption
is not sufficient for current instruments and small adjustments cause
large differences in the quality of the resulting model.  We also show
that direct observations of $T_S$ help in creating a better model
of the DGE.  Unfortunately, direct observations of $T_S$ are difficult
since they require high resolution telescopes and bright radio continuum
sources.  Some assumptions will therefore have to be made for the
regions in between bright radio sources.

It must be stated here that all of the above results are model
dependent.  The Fermi-LAT data can only provide us with the intensity of
gamma-rays from a particular direction of the sky.  Uncertainties in our
modeling of contribution other than those directly related to the H I
distribution will affect the value obtained for $T_S$.  We are currently
studying the systematic effects this will have on our results.  We also
note that even for the best fit models, the residuals show signs of
structure, strongly indicating our models are less than
perfect.


\end{document}